\documentclass[aps,prl,twocolumn,floatfix,groupedaddress,superscriptaddress,footinbib,notitlepage,showpacs]{revtex4-2}
\usepackage{graphicx,graphics,times,bm,bbm,bbold,amssymb,amsmath,amsfonts,dsfont,hyperref,color}
\usepackage[dvipsnames]{xcolor}
\usepackage{subfigure}
\hypersetup{colorlinks,linkcolor=blue,citecolor=blue,urlcolor=blue}

\newtheorem{lemma}{Lemma}
\newtheorem{theorem}{Theorem}

\usepackage[english]{babel}
\usepackage{natbib}
\usepackage[autostyle]{csquotes}
\usepackage{relsize}

\usepackage[colorinlistoftodos]{todonotes}

\newcommand{\qed}{\nobreak \ifvmode \relax \else
	\ifdim\lastskip<1.5em \hskip-\lastskip
	\hskip1.5em plus0em minus0.5em \fi \nobreak
	\vrule height0.5em width0.60em depth0.2em\fi}

\newcommand{\ignore}[1]{}

\newcommand{\re}{\mathrm{e}}

\newcommand{\average}[1]{\left\langle#1\right\rangle}
\newcommand{\ket}[1]{|#1\rangle}

\newcommand{\bra}[1]{\langle #1|}

\newcommand{\cfi}[1]{\mathpzc{F}#1}
\newcommand{\qfi}[1]{\mathbf{Q}#1}

\newcommand*{\rom}[1]{\expandafter\@slowromancap\romannumeral#1@}
\newcommand{\Tr}[1]{\mathrm{Tr}\left[#1\right]}
\let\oldsqrt\sqrt
\def\sqrt{\mathpalette\DHLhksqrt}
\def\DHLhksqrt#1#2{%
\setbox0=\hbox{$#1\oldsqrt{#2\,}$}\dimen0=\ht0
\advance\dimen0-0.2\ht0
\setbox2=\hbox{\vrule height\ht0 depth -\dimen0}%
{\box0\lower0.4pt\box2}}
\DeclareFontFamily{OT1}{pzc}{}
\DeclareFontShape{OT1}{pzc}{m}{it}%
              {<-> s * [1.25] pzcmi7t}{}
\DeclareMathAlphabet{\mathpzc}{OT1}{pzc}%
                                 {m}{it}
\hypersetup{colorlinks,linkcolor={blue},citecolor={blue},urlcolor={blue}}
\begin{document}
\title{Privacy in networks of quantum sensors}
\author{Majid Hassani}
\email{majidhasani2010@gmail.com}
\affiliation{LIP6, CNRS, Sorbonne Universit\'e, 4 place Jussieu, F-75005 Paris, France}
\author{Santiago Scheiner}
%\email{damian.markham@lip6.fr}
\affiliation{LIP6, CNRS, Sorbonne Universit\'e, 4 place Jussieu, F-75005 Paris, France}
\author{Matteo G. A. Paris}
\affiliation{Quantum Technology Lab, Universit\`a degli Studi di Milano, I-20133 Milano, Italy}
%\email{matteo.paris@fisica.unimi.it}
\author{Damian Markham}
%\email{damian.markham@lip6.fr}
\affiliation{LIP6, CNRS, Sorbonne Universit\'e, 4 place Jussieu, F-75005 Paris, France}
%%%%%%%%%%%%%%%%%%%%%%%%%%%%%%%%%%%%%%%%%%%%%%
\begin{abstract}
We treat privacy in a network of quantum sensors where accessible information is limited to specific functions of the network parameters, and all other information remains private. We develop an analysis of privacy in terms of a manipulation of the quantum Fisher information matrix, and find the optimal state achieving maximum privacy in the estimation of linear combination of the unknown parameters in a network of quantum sensors. We also discuss the effect of uncorrelated noise on the privacy of the network. Moreover, we illustrate our results with an example where the goal is to estimate the average value of the unknown parameters in the network. In this example, we also introduce the notion of quasi-privacy ($\epsilon$-privacy), quantifying how close the state is to being private.
\end{abstract}
\date{\today}
%%%%
\maketitle
%\listoftodos
%\onecolumngrid
Simultaneously estimating spatially distributed unknown parameters via a quantum network,
commonly referred to as networked quantum sensing, has a wide array of applications, 
including  clocks synchronization \cite{Komar2014,Dai2020} and phase imaging \cite{HumphreysPRL,Proctor2016,Chris2016}. Alongside experimental advancements in 
networked quantum sensing \cite{Guo2020,Liu2021}, theoretical studies are continuously 
developing to tackle the most realistic challenges in this field \cite{Proctor2017networked,ProctorPRL2018}. The inevitable presence 
of malicious adversaries eavesdropping on quantum channels is a significant hurdle 
in networked sensing. In such a case, the goal is not only to estimate unknown 
parameters with the ultimate attainable accuracy but also to ensure that the 
estimation process is done securely. Although incorporating the notions of 
security to single-parameter quantum estimation has been investigated 
\cite{Huang2014,Shettell2022,Sean2023}, it is necessary to independently 
scrutinize the concepts of security in the networked quantum sensing \cite{shettell2022private,Rahim2023}.

\begin{figure*}[ht]
    \centering
    \includegraphics[width=1.0\textwidth]{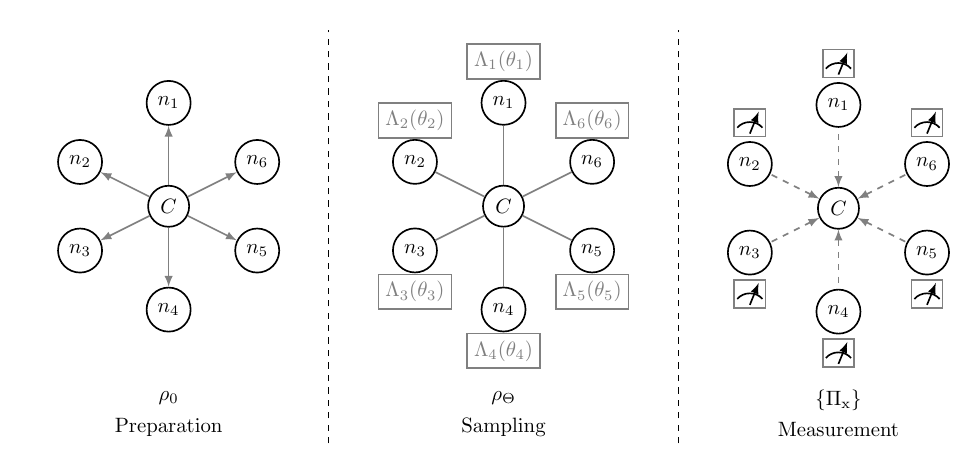}
    \caption{Schematic of a network of quantum sensors with $d=6$. After preparing and sharing the quantum probe $\rho _{0}$ by the centeral node ($C$) (preparation stage, in general it will be an entangled state), the $\mu$th unknown parameter ($\theta _{\mu}$) is encoded by local quantum operations ($\Lambda_{\mu}(\theta_{\mu})$), overall described by the factorized channel $\mathbf{\Lambda}_{\Theta}$ (sampling stage). In order to estimate the values $\Theta$, the set of parameters, the quantum probe is locally measured (measurement stage) at each node. Measurement results are sent publicly to the central node.}
    \label{Stages}
\end{figure*}

In this work, we develop the notion of \textit{privacy} introduced in \cite{shettell2022private} and its relation to standard multiparamater estimation tools, notably the quantum Fisher information matrix. The goal of a private network of quantum 
sensors is to ensure optimal precision {\em and} 
that all parties only have access to the allowed information, and not more - so that it remains private. To set the stage, let us consider a statistical model 
made of nodes, where at each node an unknown parameter $\theta _{\mu}$ 
is encoded locally on a global quantum state via a given quantum channel 
$\Lambda_{\mu} (\theta _{\mu})$. The overall channel is given by
\begin{align}\label{Dynevo}
\mathbf{\Lambda} _{\Theta}&=\bigotimes _{\mu=1}^{d} \Lambda_{\mu}(\theta _{\mu}),
\end{align}
where $\Theta=\{\theta _{1}, \theta _{2}, \cdots, \theta _{d}\}$ denotes the set 
of unknown parameters. After the encoding stage,  local measurements are performed at each node 
and the results are announced publicly. The conditional probability distribution of the 
outcomes is given by the Born rule $p(\mathrm{x}|\Theta)=\Tr{\rho _{\Theta}\Pi _{\mathrm{x}}}$ 
in which $\rho_{\Theta}$ is the quantum state of the probe after the encoding, and $\{\Pi _{\mathrm{x}}\}$ represents a 
(factorized) positive operator-valued measure (POVM) acting on the global Hilbert space 
describing the overall state at all the nodes. After collecting results $\mathrm{x}$ 
from repeated (local) measurements, one can estimate the value 
of unknown parameter $\theta_{\mu}$ by an estimator function $\tilde{\theta}_{\mu}(\mathrm{x})$. 
The general scheme of the protocol is depicted as in Fig. \ref{Stages}.

% \begin{figure}[h]
% \centering
% \includegraphics[width=0.35\textwidth]{GeneralQMetrology.pdf}
% \caption{Network of quantum sensors. After preparing the quantum probe $\rho _{0}$ (preparation stage, in general it will be an entangled state), a set of unknown parameters is encoded by local quantum operations, overall described by the factorized channel  $\mathbf{\Lambda}_{\Theta}$ (sampling stage). In order to estimate the values $\Theta$ the set of parameters, the quantum probe is locally measured (measurement stage) at each node.}
% \label{GeneralQMetrology} 
% \end{figure}

In local estimation theory, the classical Fisher information matrix (CFIm) quantifies the amount of information that may be extracted about the set of unknown parameters given the state of the probe (a.k.a. the statistical model) and a 
specific measurement. The entries of the CFIm are given by
\begin{equation}\label{CFIm}
\cfi_{\mu\nu}(\Theta)=\int\mathrm{d}\mathrm{x}~p(\mathrm{x}|\Theta)~\partial_{\mu}\ln p(\mathrm{x}|\Theta)~\partial_{\nu}\ln p(\mathrm{x}|\Theta),
\end{equation}
where $\partial_{\mu}=\frac{\partial}{\partial\theta_{\mu}}$. In turn, the CFIm determines a lower bound on the precision of estimation through the so-called multiparameter Cram{\'e}r-Rao bound \cite{Helstrom1967pla,Holevo1973,Helstrom1976,Holevo1977,Hayashi2008,Ragy2016,Demko2020,MajidMultiparameter}
\begin{equation}\label{CRB}
\text{Cov}(\Theta)\geqslant\frac{1}{\cfi{}},
\end{equation}
in which $ \text{Cov}(\Theta) $ is the $d \times d$ covariance matrix where each entry is given by
\begin{equation}\label{Covdef}
\text{Cov}_{\mu\nu}(\Theta)=\int\mathrm{d}\mathrm{x}~p(\mathrm{x}|\Theta)\left(\tilde{\theta}_{\mu}(\mathrm{x})-\theta_{\mu}\right)\left(\tilde{\theta}_{\nu}(\mathrm{x})-\theta_{\nu}\right).
\end{equation}

The metrological problem that we pose in this paper is that of estimating a global function of unknown parameters, namely $f(\Theta)$. In this setting, privacy was introduced in \cite{shettell2022private} and means that each party $\mu$ can only access $f(\Theta)$ and their own parameter $\theta_\mu$ and no other information (for example, they are not allowed to know the other parties parameters unless it is equal to $f(\Theta)$). However, that work focused on one particular function (the average of the parameters), and lacked a general way of addressing different functions. This work develops a more detailed account of privacy for any functions, which also allows a more detailed analysis of optimality and noise.

%\damian{very}{ interesting comments.}}

Such a privacy quantifier in the network of quantum sensors should capture the idea that only the information about $f(\Theta)$ can be extracted from the network of quantum sensors, but the individual values of each parameter should remain hidden.
% \begin{figure}[h]
% \centering
% \includegraphics[width=0.35\textwidth]{Fig-Network.pdf}
% \caption{Schematic of a network of quantum sensors with $d=6$. The $\mu$th unknown parameter ($\theta _{\mu}$) is encoded via a quantum channel $\Lambda _{\mu}$ at each node. The initial quantum state is distributed throughout the quantum network. Malicious adversaries may eavesdrop quantum channels to extract information. Moreover, some of the nodes can behave dishonestly and collaborate with each other to corrupt the network.}
% \label{Fig-Network} 
% \end{figure}
 
The CFIm actually depends on both the quantum statistical model $\rho_{\Theta}$ and the particular set of measurement operators $\{\Pi_{\mathrm{x}}\}$. One can set an upper bound on the CFIm, by optimizing over all possible measurements (including joint entangled measurements across the nodes). Such an upper bounds may be derived by introducing  the symmetric logarithmic derivative operator for each parameter, denoted by $L_{\mu}$, (SLD) \cite{Helstrom1976} as
\begin{equation} \label{SLD}
\partial_{\mu}\rho_{\Theta} = \frac{1}{2}\{L_{\mu}, \rho_{\Theta}\},
\end{equation}
where $\{,\}$ denotes the anticommutator. By substituting Eq. (\ref{SLD}) in Eq. (\ref{CFIm}) and employing the Cauchy-Schwarz inequality \cite{CavesBraunstein,MatteoIJQI,Watanabe2013,Yang2019}, one obtains the following upper bound on the CFIm 
\begin{equation}
\cfi _{\mu\nu}\leqslant\qfi_{\mu\nu}[\Theta],
\end{equation}
where the quantum Fisher information matrix (QFIm) is defined as 
\begin{equation}\label{QFIm}
\qfi _{\mu\nu}[\Theta]=\frac{1}{2}\Tr{\rho _{\Theta}\{L_{\mu}, L_{\nu}}\}.
\end{equation}
The QFIm is a symmetric matrix with real elements, which quantifies the maximum amount of extractable information about different unknown parameters over \textit{all} possible measurements. In particular, the off-diagonal entries of the QFIm imply that the different unknown parameters are statistically correlated to each other. If the different SLDs do no commute, the different parameters cannot be estimated independently without the addition of intrinsic noise of quantum origin.

If the aim is to estimate some function(s) of unknown parameters, $\Theta' =f(\Theta)$, the corresponding CFIm and QFIm may be obtained 
by reparametrization
\begin{align}
\cfi '&=B^{T}\cfi  B,\label{CFI-prime}\\
\qfi [\Theta']&=B^{T}\qfi[\Theta] B,\label{QFI-prime}
\end{align}
where the elements of the transformation matrix $B$ are defined as $B_{\mu\nu}=\partial \theta_{\mu}/\partial\theta ' _{\nu}$ \cite{MatteoIJQI,genoni2008optimal}.

We will now see how the notion of privacy puts constraints on the form of the QFIm, that will allow us to state conditions for privacy and lead to its quantification in our example (the average of local parameters). The starting point is to first ask that the 
reparametrized QFIm, $\qfi [\Theta']$, is a diagonal matrix. The diagonal form of the QFIm implies that there is no statistical correlation between the different linear functions of the unknown parameters (different $\theta '$s). Since the QFIm is a real symmetric positive definite matrix, it can be diagonalized by a similarity transformation. In the diagonal representation, the eigenvectors of $\qfi [\Theta]$ correspond to the coefficients of the linear combination of the unknown parameters which can be estimated in private. In particular, if the diagonal representation of $\qfi [\Theta]$ is a $1$-rank matrix, only a single linear combination of the unknown parameters can be estimated privately. This is the requirement we should impose.

Let us assume that, in fact, the aim of the network is to share an estimate of a (single) linear combination of $\Theta$; $\theta' _{1}=\mathbf{w}^{T}\Theta $ for some $\mathbf{w}\in\mathbb{R}^{d}$ \cite{Proctor2017networked,ProctorPRL2018,EldredgePRA2018}. In order to ensure privacy of this shared estimation protocol, the QFIm must be a $1$-rank matrix, i.e., $\qfi [\Theta]\propto\mathbf{w}\mathbf{w}^{T}$ (or $\qfi [\Theta]=a\mathbf{w}\mathbf{w}^{T}$ where $a$ is a real positive constant). This fact implies that the only extractable information from the network is about $\theta '_{1}$ and the local information about the parameters is kept private. For a given vector of interest like $\mathbf{w}$, one can construct $W=\mathbf{w}\mathbf{w}^{T}$. In order to get the privacy, the QFIm of the 
statistical model should be proportional to $W$.

Since the concept of privacy in quantum networks is highly sensitive to the relationships between the different entries of the QFIm, the definition of privacy can be linked to the continuity relations among them \cite{Augusiak2016,Safranek2017,Majid2019}.
Without any specific assumption about the initial states and how quantum states acquire their 
parameter dependence, we may arrive at the following Theorem which is the generalization of results in \cite{Majid2019} for the entries of the QFIm.
\begin{theorem}
Given the generic statistical model $\rho_{\Theta}$, the following inequality holds true:
\begin{widetext}
\begin{equation}
\Big\vert\qfi _{\mu\nu}[\Theta]-\qfi _{\mu'\nu'}[\Theta]\Big\vert \leqslant \frac{1}{2}\xi\, \Big[\Vert\partial _{\mu}\rho _{\Theta}-\partial _{\mu'}\rho _{\Theta}\Vert_{1}\left(\Vert\partial _{\nu}\rho _{\Theta}\Vert_{1}+\Vert\partial _{\nu'}\rho _{\Theta}\Vert_{1}\right)+\Vert\partial _{\nu}\rho _{\Theta}-\partial _{\nu'}\rho _{\Theta}\Vert_{1}\left(\Vert\partial _{\mu}\rho _{\Theta}\Vert_{1}+\Vert\partial _{\mu'}\rho _{\Theta}\Vert_{1}\right)\Big],\label{continuityQFIm-entry}
\end{equation}
\end{widetext}
where 
\begin{equation}\label{xi}
\xi=\frac{1}{\lambda _{\text{min}}(\tilde{\rho})}\left(1+\frac{32}{\lambda _{\text{min}}(\tilde{\rho})}\right),
\end{equation}
and $\tilde{\rho}$ is the (invertible) restriction of $\rho$ onto the support subspace of the quantum state.
\end{theorem}
\textit{Proof:} See Appendix \ref{proof of continuity relation} for the complete proof.
$\hfill\blacksquare$

Such a continuity relation not only can help to find a proper initial state which provides privacy in the networked sensing but also paves the way to define quasi-privacy or $\epsilon$-privacy, which will be considered later in this letter. 

In order to obtain better insight about the applications of the above results, let us consider the case where $\mathbf{w}^{T}=(\omega _{1}, \omega _{2},\cdots, \omega _{d}),~\forall\omega _{\mu}\in\mathbb{R}$ . This yields
\begin{align}\label{W-linear}
W&=\mathbf{w}\mathbf{w}^{T}\nonumber\\
&=\left(\begin{array}{cccc}
\omega _{1}\omega _{1}&\omega _{1}\omega _{2}&\cdots&\omega _{1}\omega _{d}\\ 
\omega _{2}\omega _{1}&\omega _{2}\omega _{2}&\cdots&\omega _{2}\omega _{d}\\
\vdots&\vdots&\ddots&\vdots\\
\omega _{d}\omega _{1}&\omega _{d}\omega _{2}&\cdots&\omega _{d}\omega _{d}\\
\end{array}\right).
\end{align}
To obtain the privacy in the estimation of $\theta' _{1}=\mathbf{w}^{T}\Theta $, the QFIm should be proportional to $W$,
\begin{equation}\label{PriConLin}
  \qfi _{\mu\nu}[\Theta]\propto W_{\mu\nu}\Rightarrow\qfi _{\mu\nu}[\Theta]\propto \omega_{\mu}\omega_{\nu},~~~\forall\mu, \nu.  
\end{equation}
For the purpose of finding proper quantum states where their corresponding QFIm satisfy Eq. (\ref{PriConLin}), the continuity relation, Eq. (\ref{continuityQFIm-entry}), can be recast as follows 
\begin{equation}\label{Con-priv-Check}
    \Big\vert\qfi _{\mu\mu}[\Theta]-\qfi _{\mu\nu}[\Theta]\Big\vert \leqslant \xi'\Vert\partial _{\mu}\rho _{\Theta}-\partial _{\nu}\rho _{\Theta}\Vert_{1},~~\forall\mu\neq\nu,
\end{equation}
where $\xi'$ includes all other terms that are not pertinent to the rest of the derivation. Substituting Eq. (\ref{PriConLin}) in Eq. (\ref{Con-priv-Check}), gives
\begin{equation}\label{Con-priv-Check-2}
    \vert\omega _{\mu}-\omega _{\nu}\vert \leqslant \zeta\Vert\partial _{\mu}\rho _{\Theta}-\partial _{\nu}\rho _{\Theta}\Vert_{1},~~\forall\mu\neq\nu,
\end{equation}
in which $\zeta=\xi'/\vert\omega_{\mu}\vert$. Since the proportionality is crucial here, without loss of generality, Eq. (\ref{Con-priv-Check-2}) can be rephrased as follows 
\begin{equation}\label{Priv-Condition-linear}
    \Vert\partial _{\mu}\rho _{\Theta}-\partial _{\nu}\rho _{\Theta}\Vert_{1}\propto\vert\omega _{\mu}-\omega _{\nu}\vert,~~\forall\mu\neq\nu.
\end{equation}
Hence, any quantum state which satisfies the above condition (Eq. (\ref{Priv-Condition-linear}))
can estimate $\theta' _{1}$ in private  irrespective of how acquires the parameter dependence. In the following, we specify our study to the case where the unknown parameters are encoded via local unitary evolutions, $U(\theta _{\mu})=\re ^{-iH_{\mu}(\theta _{\mu})}$ onto a shared quantum state. Here $H_{\mu}(\theta _{\mu})$ is a Hermitian operator that acts non-trivially on the Hilbert space of each quantum sensor. Hence, the sampling operator can be presented by
\begin{align}\label{Unievo}
\mathbf{U}_{\Theta}&=\bigotimes _{\mu=1}^{d} U(\theta _{\mu})\nonumber\\
&=\re ^{-i\sum _{\mu}\mathrm{H}_{\mu}},
\end{align}
where $\mathrm{H}_{\mu}=\mathbb{1}\otimes\mathbb{1}\otimes\cdots\otimes (H_{\mu}(\theta _{\mu}))^{\otimes\omega_{\mu}}\otimes\cdots\otimes\mathbb{1}\otimes\mathbb{1}$. The first derivative of the density matrix in the case of unitary evolution is derived as follows
\begin{equation}\label{first-der-rho}
\partial _{\mu}\rho_{\Theta}=-i[\mathrm{H}_{\mu}^{'},\rho_{\Theta}],
\end{equation}
where $[,]$ denotes the commutator and $\mathrm{H}_{\mu}^{'}=\partial _{\mu}\mathrm{H}_{\mu}$. From whence the condition (\ref{Priv-Condition-linear}) can be cast in this form 
\begin{equation}\label{condition-linear-Ham}
\Vert[\mathrm{H}_{\mu}^{'}-\mathrm{H}_{\nu}^{'},\rho _{\Theta}]\Vert_{1}\propto\vert\omega _{\mu}-\omega_{\nu}\vert~~~\forall \mu\neq\nu.
\end{equation}
For the unitary evolutions where their associated generators satisfy 
\begin{equation}\label{comm-generators}
[\partial _{\mu}H_{\mu}(\theta _{\mu}),H_{\mu}(\theta _{\mu})]=0~~~\forall \mu,
\end{equation}
Eq. (\ref{condition-linear-Ham}) can be simplified more. Using the fact that $\rho_{\Theta}=\mathbf{U}_{\Theta}\rho_{0}\mathbf{U}_{\Theta}^{\dagger}$ and 
\begin{equation}
[\mathpzc{A},\mathpzc{B}\mathpzc{C}\mathpzc{D}]=[\mathpzc{A},\mathpzc{B}]\mathpzc{C}\mathpzc{D}+\mathpzc{B}\mathpzc{C}[\mathpzc{A},\mathpzc{D}]+\mathpzc{B}[\mathpzc{A},\mathpzc{C}]\mathpzc{D},
\end{equation}
for any arbitrary operators $\mathpzc{A}, \mathpzc{B}, \mathpzc{C},$ and $\mathpzc{D}$, Eq. (\ref{condition-linear-Ham}) yields
\begin{equation}\label{condition-linear-Ham-0}
\Vert[\mathrm{H}_{\mu}^{'}-\mathrm{H}_{\nu}^{'},\rho _{0}]\Vert_{1}\propto\vert\omega _{\mu}-\omega_{\nu}\vert~~~\forall \mu\neq\nu.
\end{equation}
In order to estimate the linear combination of spatially distributed unknown parameters (which are encoded via local unitary operations where their generators satisfy Eq. (\ref{comm-generators}), the initial state of quantum probe should satisfy Eq. (\ref{condition-linear-Ham-0}). Let us consider the case of multiplicative unknown parameter in which $H_{\mu}(\theta _{\mu})=\theta _{\mu}H$ (where satisfies Eq. (\ref{comm-generators}) ). Ergo
\begin{equation}\label{derivH-multipli}
\mathrm{H}'_{\mu}=\mathbb{1}\otimes\mathbb{1}\otimes\cdots\otimes (\omega_{\mu}H\otimes(\theta _{\mu}H)^{\otimes\omega_{\mu}-1})\otimes\cdots\otimes\mathbb{1}\otimes\mathbb{1}.
\end{equation}
In this case any pure states in the form of 
\begin{equation}
\ket{\Psi}=\sum_{i=1}^{n}\alpha_{i}\bigotimes_{\mu=1}^{d}\ket{\lambda _{i}}^{\otimes \omega_{\mu}},\label{Global-GHZ}
\end{equation}
where $\alpha _{i}\in\mathbb{C}$ and $\{\ket{\lambda _{i}}\}$ are the eigenvectors of $n$-dimensional $H$, satisfy condition (\ref{condition-linear-Ham-0}) and provide privacy in the estimation of the linear combination with integer coefficients in the networked sensing.

\textit{Noise model.}---We now analyse the effect of noise. Generally, noise can affect any metrological schemes after or before the sampling stage. %Fig. (\ref{NoisyQMetrology}). 
%\begin{figure}[h]
%\centering
%\includegraphics[width=0.35\textwidth]{NoisyQMetrology.pdf}
%\caption{Metrological scheme can be corrupted by various noise models. These noise models can occur: (a) after the sampling stage or (b) before the sampling stage.}
%\label{NoisyQMetrology} 
%\end{figure}
%\santiago{Figure 2?}{Figure \ref{NoisyQMetrology} can be removed now, maybe.}
Let us consider the case where the quantum probe satisfies condition (\ref{Priv-Condition-linear}) and the noise affects the probe state after the sampling stage, %Fig. (\ref{NoisyQMetrology}, (a))
\begin{align}
\rho'_{\Theta}=\mathbf{\Lambda} _{\Theta}(\rho _{0})&=\sum_{\mathbf{k}=1}^{q^{d}}\mathbf{A}_{\mathbf{k}}\mathbf{U}(\Theta)\rho _{0}\mathbf{U}(\Theta)^{\dagger}\mathbf{A}_{\mathbf{k}}^{\dagger}=\sum_{\mathbf{k}}\mathbf{A}_{\mathbf{k}}\rho _{\Theta}\mathbf{A}_{\mathbf{k}}^{\dagger},
\end{align}
where $\mathbf{A}_{\mathbf{k}}=A_{k_{1}}\otimes A_{k_{2}}\otimes\cdots \otimes A_{k_{d}}$ in which $\mathbf{k}=\{k_{1}, k_{2}, \cdots, k_{d}\}$. In this notation $k_{i}\in\{1,2,\dots,q\}$ denotes the $k_{i}$th Kraus operator of the noise model which satisfies $\sum_{k=1}^{q}A_{k}^{\dagger}A_{k}=\mathbb{1}$ and acts on the $i$th node of the network \cite{Fujiwara2008}. Without loss of generality, one can consider the case where the Kraus operators do not depend on the set of unknown parameters. Hence,
\begin{align}\label{noise-aftersampling}
\Vert\partial _{\mu}\rho' _{\Theta}-\partial _{\nu}\rho' _{\Theta}\Vert_{1}&=\big\Vert\sum_{\mathbf{k}=1}^{q^{d}}\mathbf{A}_{\mathbf{k}}(\partial _{\mu}\rho _{\Theta}-\partial _{\nu}\rho _{\Theta})\mathbf{A}_{\mathbf{k}}^{\dagger}\big\Vert_{1}\nonumber\\
&\propto\vert\omega _{\mu}-\omega _{\nu}\vert,~~\forall\mu\neq\nu,
\end{align}
which shows that the probe state remains private. 

We explore the privacy for the cases in which the noise affects the quantum probe among the preparation stage and the sampling stage. Let us suppose the quantum states which provide the privacy in the ideal case, have been shared throughout the network. If all Kraus operators of the noise model commute with the sampling operators, one can still estimate the parameter of interest in private. Since we can separate the noise model and the sampling operators, the same approach like relation (\ref{noise-aftersampling}) holds true. 
%following example. Consider the case where the aim is to estimate the average value of the spatially distributed unknown parameters which are encoded via local evolutions, Eq. (\ref{Dynevo}). In this case our parameter of interest is $\bar{\theta}=\mathbf{w}^{T}\Theta$ where $\mathbf{w}^{T}=1/d~(1, 1,\cdots, 1)$. From whence
%\begin{align}\label{W-average}
%W&=\mathbf{w}\mathbf{w}^{T}\nonumber\\
%1&1&\cdots&1\\ 
%1&1&\cdots&1\\
%\vdots&\vdots&\ddots&\vdots\\
%1&1&\cdots&1\\
%\end{array}\right).
%\end{align}

\textit{Example.}---We now consider the specific case in which the aim is to estimate the average value of the spatially distributed unknown parameters which are encoded via local evolutions, Eq. (\ref{Dynevo}). In this case our parameter of interest is $\bar{\theta}=\mathbf{w}^{T}\Theta$ where $\mathbf{w}^{T}=1/d~(1, 1,\cdots, 1)$. Hence, $\vert\omega_{\mu}-\omega _{\nu}\vert=0, \forall\mu,\nu$. This implies that all entries of the QFIm should be equal to each other. From Eq. (\ref{Priv-Condition-linear}), if all first derivatives of the probe state (after the sampling stage) with respect to the different unknown parameters are equal, then all entries of the QFIm are equal to each other. Thus any quantum states which satisfy the following condition
\begin{equation}\label{condition-average}
\partial _{\mu}\rho_{\Theta}=\partial _{\nu}\rho_{\Theta}~~~~~~\forall \mu,\nu,
\end{equation}
can be used in the private estimation of the average value irrespective of how acquires the parameter dependence. Once more, we can consider the case of unitary evolution with multiplicative unknown parameter where $H=\sigma _{z}/2$
Therefore, the unitary evolution reads
\begin{align}\label{Unitary-ave}
\mathbf{U}(\Theta)&=\bigotimes _{\mu=1}^{d}U(\theta _{\mu})\nonumber\\
&=\bigotimes _{\mu=1}^{d}(\ket{0}\bra{0}+\re ^{-i\theta _{\mu}}\ket{1}\bra{1}).
\end{align}
From whence, the privacy condition in Eq. (\ref{condition-average}) can be written as
\begin{equation}\label{condition-average-uni}
[\mathrm{H}_{\mu}^{'}-\mathrm{H}_{\nu}^{'},\rho _{\Theta}]=0~~~~~~\forall \mu,\nu.
\end{equation}
Now, by substituting the eigenvectors of $\sigma _{z}$ in Eq. (\ref{Global-GHZ}), one can find the private states in the form of 
\begin{equation}
\ket{\Phi}=\alpha\ket{0}^{\otimes d}+\beta\ket{1}^{\otimes d}\equiv\ket{ {\text{\footnotesize{GHZ-like}}}},\label{Gen-GHZ}
\end{equation}
where $\alpha ^{2}+\beta ^{2}=1$ (can be named as GHZ-like state) or mixed states like 
\begin{equation}
\gamma _{0}\ket{\Phi}\bra{\Phi}+\sum _{i}{\gamma _{i}\ket{\phi _{i}}\bra{\phi _{i}}},\label{mixed-priv-ave}
\end{equation}
where $\ket{\phi _{i}}=\ket{l_{1}, l_{2}, \cdots, l_{d}}$, $l_{j}\in\{0,1\}$, and $\sum _{i=0}\gamma _{i}=1$. Such states satisfy condition (\ref{condition-average-uni}) and get the privacy in the estimation of the average value. Practically speaking, one can distribute a GHZ state, $\ket{\psi _{0}}=\frac{1}{\sqrt{2}}(\ket{0}^{\otimes d}+\ket{1}^{\otimes d})$ throughout the network. Each node encodes the unknown parameter on the shared state. Hence the quantum state of the probe is given by
\begin{equation}
\ket{\psi _{\Theta}}=\frac{1}{\sqrt{2}}(\ket{0}^{\otimes d}+\re ^{-id\bar{\theta}}\ket{1}^{\otimes d}).\label{GHZ-Ave}
\end{equation}

From Eq. (\ref{GHZ-Ave}) it is obvious that the only extractable information is the average value. Ergo, we can ask each node to perform the measurement in the $X$ basis and announce the result in the public \cite{Komar2014}, Fig. \ref{Stages}. Regarding the result of the measurement, the conditional probability distribution can be derived as
\begin{equation}
p(\pm |\Theta)=\frac{1\pm\cos(d\bar{\theta})}{2^{d}},\label{conprobGHZ}
\end{equation}
where $\pm$ represents the result of the parity measurement. One can easily calculate the entries of the CFIm (Eq. (\ref{CFIm})) for the given conditional probability distribution in Eq. (\ref{conprobGHZ}) as follows
\begin{align}
\cfi_{\mu\mu}(\Theta)=1,~~~~~~~~~~~~~\cfi_{\mu\nu}(\Theta)=1,~~~\text{for}~~\mu\neq\nu .
\end{align}
This form of the CFIm implies that the information about all unknown parameters is distributed equally throughout the network. One can also calculate the QFIm by exploiting the fact that for pure states $ \left( \varrho^{2}=\varrho =\ket{\Psi }\bra{\Psi }\right) $, $ L_{\mu}=2\partial _{\mu}\varrho=2\left( \ket{\partial _{\mu}\Psi }\bra{\Psi }+\ket{\Psi }\bra{\partial_{\mu}\Psi }\right)  $. The elements of the QFIm are given by 
\begin{align}
\qfi _{\mu\nu}[\ket{\Psi }\bra{\Psi }]&=4\,\mathfrak{R}\left(\average{\partial _{\mu}\Psi\Big\vert\partial _{\nu}\Psi}-\average{\partial _{\mu}\Psi\Big\vert\Psi}\average{\Psi\Big\vert\partial _{\nu}\Psi} \right)\label{qfi-pure},
\end{align}
where $ \mathfrak{R} $ denotes the real part. Substituting Eq. (\ref{GHZ-Ave}) in Eq. (\ref{qfi-pure}), yields
\begin{align}\label{qfi-GHZ}
\qfi _{\mu\nu}[\ket{\psi _{\Theta} }\bra{\psi _{\Theta}}]&=\left(\begin{array}{cccc}
1&1&\cdots&1\\ 
1&1&\cdots&1\\
\vdots&\vdots&\ddots&\vdots\\
1&1&\cdots&1\\
\end{array}\right).
\end{align}
Since the QFIm in Eq. (\ref{qfi-GHZ}) is proportional to the $W$ matrix, the GHZ state is the appropriate initial state to estimate the average value in the network of quantum sensors privately. Since any quantum state in the form of $\varrho=\ket{\Psi}\bra{\Psi}$ is a private state (in the estimation of average function), we can define $\epsilon$-privacy in the sense of the closeness of an arbitrary state to the ideal state which provides the (perfect) privacy, e.g. $\varrho$. Given $\sigma$, the $\epsilon$-privacy may be quantified
\begin{widetext}
\begin{equation}\label{epsilonprivacy}
\epsilon=\Vert[\mathrm{H}_{\mu}^{'}-\mathrm{H}_{\nu}^{'},\sigma]\Vert_{1}=\Vert[\mathrm{H}_{\mu}^{'}-\mathrm{H}_{\nu}^{'},\sigma -\varrho]\Vert _{1}\leqslant 4\Vert \mathrm{H}_{\mu}^{'}\Vert _{\infty}\Vert\sigma -\varrho\Vert _{1}\leqslant 4\Vert H\Vert _{\infty}\Vert\sigma -\varrho\Vert _{1}\leqslant 8\Vert H\Vert _{\infty}\sqrt{1-F^{2}(\sigma ,\varrho)},
\end{equation}
\end{widetext}
where $F(\sigma ,\varrho)$ denotes the fidelity of two quantum states $F(\sigma ,\varrho)=\Tr{\sqrt{\sqrt{\varrho}\sigma\sqrt{\varrho}}~}$. The last inequality follows from $1-F(\sigma ,\varrho)\leqslant\frac{1}{2}\Vert\sigma -\varrho\Vert_{1}\leqslant\sqrt{1-F^{2}(\sigma ,\varrho)}$. Eq. (\ref{epsilonprivacy}) shows that the privacy of the network is a continuous function of fidelity, which in turn implies the robustness of our protocol against noise. In other words, some form of privacy may be achieved also for suboptimal states in a neighborhood of the optimal one.

In the mentioned example where the sampling operator is given by Eq. (\ref{Unitary-ave}), the corresponding Kraus operators of dephasing and erasure noise commute with the unitary $U(\theta _{\mu})$. The canonical Kraus operators of dephasing noise are
\begin{equation}\label{Kraus-dephasing}
A_{1}=\sqrt{1-\eta}\mathbb{1},~~~~~~~A_{2}=\sqrt{\eta}\sigma_{z},
\end{equation}
where $\mathbb{1}=\ket{0}\bra{0}+\ket{1}\bra{1}$, and $0\leqslant\eta\leqslant1$ denotes the dephasing parameter. The erasure noise is the effective way to model \textit{loss} in an optical interferometry. The erasure noise can be described as a quantum channel where transforms $\rho\mapsto(1-\eta)\rho+\eta\ket{e}\bra{e}$ which means the probe does not change with probability $1-\eta$ while with probability $\eta$ its state changes to the quantum state, $\ket{e}\bra{e}$, which is in the orthogonal subspace where the sampling takes place \cite{KolodynskyNatComm,RafalPO,Demkowicz-Dobrzanski2014}. In order to obtain the Kraus operators of this noise model, we need to add the third dimension corresponds to $\ket{e}\bra{e}$ where the sampling operator does not apply there. Hence, the Kraus operators of erasure noise are given by
\begin{align}\label{Kraus-erasure}
A_{1}&=\left(\begin{array}{ccc}
\sqrt{1-\eta}&0&0\\ 
0&\sqrt{1-\eta}&0\\
0&0&0\\
\end{array}\right),~~A_{2}=\left(\begin{array}{ccc}
0&0&0\\ 
0&0&0\\
0&0&1\\
\end{array}\right),\nonumber\\
A_{3}&=\left(\begin{array}{ccc}
0&0&0\\ 
0&0&0\\
\sqrt{\eta}&0&0\\
\end{array}\right),~~~~~~~~~~~~~~~~~~~~A_{4}=\left(\begin{array}{ccc}
0&0&0\\ 
0&0&0\\
0&\sqrt{\eta}&0\\
\end{array}\right),
\end{align}
One can easily investigate that the Kraus operators of dephasing noise, Eq. (\ref{Kraus-dephasing}), and the Kraus operators of erasure noise, Eq. (\ref{Kraus-erasure}), commute with the unitary $U(\theta _{\mu})$, Eq. (\ref{Unitary-ave}). From whence, for any initial state, like Eqs. (\ref{Gen-GHZ}) and (\ref{mixed-priv-ave}), which satisfies the condition (\ref{condition-average}), affecting the dephasing noise and the erasure noise before or after the sampling stage maintain the privacy.

However, if the Kraus operators of the noise model do not commute with the sampling operators, the presence of noise before the sampling stage can generally affect privacy. For example, in our case where the sampling stage is presented by Eq. (\ref{Unitary-ave}), the Kraus operators of the quantum depolarizing noise and amplitude damping noise do not commute with the unitary evolution. The depolarizing noise can be considered as a quantum channel where transforms $\rho\mapsto(1-\eta)\rho+\frac{\eta}{2}\mathbb{1}$, which means with the probability $1-\eta$ the quantum state remains fixed while with the probability $\eta$ the quantum state changes to the maximally mixed state. The Kraus operators of this channel are given by
\begin{equation}\label{Kraus-depolarizing}
A_{1}=\sqrt{\frac{\eta}{4}}\sigma _{x},A_{2}=\sqrt{\frac{\eta}{4}}\sigma _{y},A_{3}=\sqrt{\frac{\eta}{4}}\sigma _{z},A_{4}=\sqrt{1-\frac{3\eta}{4}}\mathbb{1}.
\end{equation}
The Kraus operators which describe the amplitude damping noise are presented as
\begin{align}\label{Kraus-amplitudedam}
A_{1}&=\left(\begin{array}{cc}
1&0\\ 
0&\sqrt{1-\eta}\\
\end{array}\right),~~~~~~~~~A_{2}=\left(\begin{array}{cc}
0&\sqrt{\eta}\\ 
0&0\\
\end{array}\right).
\end{align}
Clearly the Kraus operators of both channels (Eqs. (\ref{Kraus-depolarizing}) and (\ref{Kraus-amplitudedam})) do not commute with the sampling operators (Eq. (\ref{Unitary-ave})). Therefore, relation (\ref{noise-aftersampling}) no longer holds true. Despite of the fact, the initial GHZ-like states (Eq. (\ref{Gen-GHZ})) still preserve privacy in the presence of the depolarizing and amplitude damping noise. The straightforward proof of the privacy robustness of the initial GHZ-like states against depolarizing and amplitude damping noise is as follows. One can consider $\rho _{\text{\tiny{GHZ-like}}}=\ket{\Phi}\bra{\Phi}$ as a pure initial state. In the presence of depolarizing noise, the state of probe before the sampling stage %(Fig. (\ref{NoisyQMetrology}, (b))) 
reads 
\begin{equation}\label{effe-noise-depol}
(1-\eta)\rho _{\text{\tiny{GHZ-like}}}+\frac{\eta}{2}\mathbb{1},
\end{equation} 

which explicitly satisfies condition (\ref{condition-average}). In the case of amplitude damping noise, the GHZ-like quantum probe state changes to
\begin{equation}\label{effe-noise-amplitude}
a\rho _{\text{\tiny{GHZ-like}}}(\eta)+b\rho_{\text{\tiny{diagonal}}}(\eta),
\end{equation} 
where $a$ and $b$ are two arbitrary coefficients ($a+b=1$)---see Appendix \ref{deriv-Eq-effe-noise-amplitude} for derivation. Regarding the diagonal form of $\mathrm{H}_{\mu}^{'}$  while $H_{\mu}=\sigma _{z}/2$, Eq. (\ref{effe-noise-amplitude}) satisfies condition (\ref{condition-average}) which shows the privacy robustness of the GHZ-like state against amplitude damping noise. 

\textit{Conclusion.}---We have given a quantitative definition of privacy in the estimation of linear combination of unknown parameters where are spatially distributed in a network, in the sense that specific information can be extracted from the network of quantum sensors. Regarding the function (linear combination of unknown parameters) of interest to be estimated and the continuity relation between different entries of the QFIm, one can find the proper initial state which estimate the function privately. The effect of uncorrelated noise in the private estimation has been studied.

\textit{Acknowledgement.}--- S.S acknowledges the PEPR integrated project EPiQ ANR-22-PETQ-0007 part of Plan France 2030 and QIA, which has received funding from the European Union's Horizon 2020 research and innovation programme under grant agreement No 820445 and from the Horizon Europe grant agreements 101080128 and 101102140. MGAP acknowledges support from Italian Ministry of Research and Next Generation EU via the PRIN 2022 project RISQUE (contract n. 2022T25TR3).

%%%%%%%%%%%%%%%%%%%%%%%%%%%%%%%%%%%%%%%%%%%%%%%%%%%%%%
\vfill \pagebreak \onecolumngrid
%%%%%%%%%%%%%%%%%%%%%%%%%%%%%%%%%%%%%%%%%%%%%%%%%%%%%%%%%%%%%%%%
\appendix
\section{Derivation of the continuity relation; Equation (\ref{continuityQFIm-entry})}\label{proof of continuity relation}
Here we present the derivation of the continuity relation (Eq. (\ref{continuityQFIm-entry})). We begin by the alternative relation of the QFIm
\begin{align}
\qfi _{\mu\nu}[\Theta]&=\frac{1}{2}\Tr{\rho _{\Theta}\lbrace L_{\mu},L_{\nu}\rbrace}\nonumber\\
&=\frac{1}{2}\left( \frac{1}{2}\Tr{\rho _{\Theta} L_{\mu}L_{\nu}}+\frac{1}{2}\Tr{\rho _{\Theta} L_{\mu}L_{\nu}}+\frac{1}{2}\Tr{\rho _{\Theta} L_{\nu}L_{\mu}}+\frac{1}{2}\Tr{\rho _{\Theta} L_{\nu}L_{\mu}}\right) \nonumber\\
&=\frac{1}{2}\left( \Tr{\left(\frac{\rho _{\Theta} L_{\mu}+L_{\mu}\rho _{\Theta}}{2}\right)L_{\nu}}+\Tr{\left(\frac{\rho _{\Theta} L_{\nu}+L_{\nu}\rho _{\Theta}}{2}\right)L_{\mu}}\right) \nonumber\\
&\overset{(\ref{SLD})}{=}\frac{1}{2}\Tr{\partial _{\mu}\rho _{\Theta}\, L_{\nu}+\partial _{\nu}\rho _{\Theta} \, L_{\mu}}.\label{alter-def-QFI}
\end{align}
The difference between two arbitrary entries of the QFIm is given by
\begin{align}
\vert\qfi _{\mu\nu}[\Theta]-\qfi _{\mu'\nu'}[\Theta]\vert&=\frac{1}{2}\left(\Tr{\partial _{\mu}\rho _{\Theta}\, L_{\nu}+\partial _{\nu}\rho _{\Theta} \, L_{\mu}}-\Tr{\partial _{\mu'}\rho _{\Theta}\, L_{\nu'}+\partial _{\nu'}\rho _{\Theta} \, L_{\mu'}}\right)\nonumber\\
&=\frac{1}{2}\left(\Tr{\partial _{\mu}\rho _{\Theta}\, L_{\nu}-\partial _{\mu'}\rho _{\Theta}\, L_{\nu'}}+\Tr{\partial _{\nu}\rho _{\Theta} \, L_{\mu}-\partial _{\nu'}\rho _{\Theta} \, L_{\mu'}}\right) \nonumber\\
&=\frac{1}{2}\left(\Tr{\partial _{\mu}\rho _{\Theta}\, (L_{\nu}-L_{\nu'})+(\partial _{\mu}\rho _{\Theta}-\partial _{\mu'}\rho _{\Theta})\, L_{\nu'}}+\Tr{\partial _{\nu}\rho _{\Theta}\, (L_{\mu}-L_{\mu'})+(\partial _{\nu}\rho _{\Theta}-\partial _{\nu'}\rho _{\Theta})\, L_{\mu'}}\right),\label{diff-QFI}
\end{align}
where in the last line, we have used the fact that \cite{Majid2019}
\begin{align*}
\mathpzc{A}\mathpzc{B}-\mathpzc{A'}\mathpzc{B'}&=\mathpzc{A}\mathpzc{B}-\mathpzc{A}\mathpzc{B'}+\mathpzc{A}\mathpzc{B'}-\mathpzc{A'}\mathpzc{B'},\nonumber\\
&=\mathpzc{A}(\mathpzc{B}-\mathpzc{B'})+(\mathpzc{A}-\mathpzc{A'})\mathpzc{B'}.
\end{align*}
In order to derive an upper bound on Eq. (\ref{diff-QFI}), we apply the same approach as Ref. \cite{Majid2019}
\begin{align}
\vert\qfi _{\mu\nu}[\Theta]-\qfi _{\mu'\nu'}[\Theta]\vert\leqslant\frac{1}{2}\left(\Vert\partial _{\mu}\rho _{\Theta}\Vert _{1}\Vert L_{\nu}-L_{\nu'}\Vert_{\infty}+\Vert\partial _{\mu}\rho _{\Theta}-\partial _{\mu'}\rho _{\Theta}\Vert _{1}\Vert L_{\nu'}\Vert_{\infty}+\Vert\partial _{\nu}\rho _{\Theta}\Vert _{1}\Vert L_{\mu}-L_{\mu'}\Vert_{\infty}+\Vert\partial _{\nu}\rho _{\Theta}-\partial _{\nu'}\rho _{\Theta}\Vert _{1}\Vert L_{\mu'}\Vert_{\infty}\right),\label{con-QFI-1}
\end{align}
in which the relation $\vert\Tr{\mathpzc{B}\mathpzc{A}}\vert\leqslant\Vert\mathpzc{A}\Vert_{1}\Vert\mathpzc{B}^{\dagger}\Vert_{\infty}$ has been utilized. Directly from Ref. \cite{Majid2019}, we know
\begin{align}
\Vert L_{\mu}\Vert_{\infty}&\leqslant\xi \Vert\partial _{\mu}\rho _{\Theta}\Vert_{1},\label{def-normL}\\
\Vert L_{\mu}-L_{\nu}\Vert_{\infty}&\leqslant\xi \Vert\partial_{\mu}\rho _{\Theta}-\partial_{\nu}\rho _{\Theta}\Vert_{1},\label{def-normL-diff}
\end{align}
where
\begin{equation*}
\xi=\frac{1}{\lambda _{\text{min}}(\tilde{\rho})}\left(1+\frac{32}{\lambda _{\text{min}}(\tilde{\rho})}\right).
\end{equation*}
Substituting Eqs. (\ref{def-normL}) and (\ref{def-normL-diff}) in Eq. (\ref{con-QFI-1}) yields
\begin{equation*}
\left\vert\qfi _{\mu\nu}[\Theta]-\qfi _{\mu'\nu'}[\Theta]\right\vert\leqslant\frac{1}{2}\xi\big[\Vert\partial _{\mu}\rho _{\Theta}-\partial _{\mu'}\rho _{\Theta}\Vert_{1}\left(\Vert\partial _{\nu}\rho _{\Theta}\Vert_{1}+\Vert\partial _{\nu'}\rho _{\Theta}\Vert_{1}\right)+\Vert\partial _{\nu}\rho _{\Theta}-\partial _{\nu'}\rho _{\Theta}\Vert_{1}\left(\Vert\partial _{\mu}\rho _{\Theta}\Vert_{1}+\Vert\partial _{\mu'}\rho _{\Theta}\Vert_{1}\right)\big].
\end{equation*}
%%%%%%%%%%%%%%%%%%%%%%%%%%%%%%%%%%%%%%%%%%%%%%%%%%%%%%%%%%%%%%%%
\section{Derivation of Equation (\ref{effe-noise-amplitude})}\label{deriv-Eq-effe-noise-amplitude}
Here we show that the final state of the GHZ-like quantum probe state (Eq. (\ref{Gen-GHZ})) after the effect of amplitude damping noise (Eq. (\ref{Kraus-amplitudedam})) is proportional to 
\begin{equation}\label{effe-noise-amplitude-app}
\rho _{\text{\tiny{GHZ-like}}}(\eta)+\rho_{\text{\tiny{diagonal}}}(\eta).
\end{equation}
The Kraus operators of amplitude damping noise read
\begin{align}\label{Kraus-amplitudedam-app}
A_{1}&=\left(\begin{array}{cc}
1&0\\ 
0&\sqrt{1-\eta}\\
\end{array}\right),~~~~~~~~~A_{2}=\left(\begin{array}{cc}
0&\sqrt{\eta}\\ 
0&0\\
\end{array}\right).
\end{align}
Both Kraus operators $A_{1}$ and $A_{2}$ (Eq. (\ref{Kraus-amplitudedam-app})) have the following properties:
\begin{itemize}
\item Generalized permutation (GP) matrix : a matrix has at most one non-zero entry in each row and each column.
\item Upper triangular matrix: a square matrix whose all entries below the diagonal are zero.
\end{itemize}
In our case of interest (Eq. (\ref{Kraus-amplitudedam-app})), we relaxed the invertibility of the permutation matrix by considering that there exists at most one non-zero entry in each row and each column. In the following we present auxiliary lemmas and corollary.
\begin{lemma}\label{Lemma1}
The tensor product of two GP matrices is a GP matrix.
\end{lemma}
\textit{Proof:} Let $A, B \in \mathbb{C}^{n\times n}$ be two arbitrary GP matrices. From whence
\begin{align}\label{lemma1-C}
C&=A\otimes B\nonumber\\
&=\left(\begin{array}{ccccc}
a_{11}B&a_{12}B&\cdots&a_{1n}B\\ 
\vdots&\vdots&&\vdots\\ 
a_{i1}B&a_{i2}B&\cdots&a_{in}B\\
\vdots&\vdots&&\vdots\\ 
a_{n1}B&a_{n2}B&\cdots&a_{nn}B\\
\end{array}\right).
\end{align}
Let consider $a_{ij}B$ as an arbitrary block of $C\in\mathbb{C}^{n^{2}\times n^{2}}$. If $a_{ij}\neq 0$, then all other blocks like $a_{ik}B$ ($\forall k\neq j$) and $a_{k'j}B$ ($\forall k'\neq i$) are equal to zero. It means that $a_{ij}B$ is the only non-zero block in $i$th row and $j$th column. Since $B$ again is a GP matrix, then $a_{ij}B$ block is also a GP matrix.
$\hfill\blacksquare$\\
\textbf{Corollary 1} Since $A_{1}$ and $A_{2}$ are two GP matrices, $\mathbf{A}_{\mathbf{k}}=A_{k_{1}}\otimes A_{k_{2}}\otimes\cdots\otimes A_{k_{d}}$ is a GP matrix $\forall~\mathbf{k}\in\{{k_{1}, k_{2}, \cdots, k_{d}}\}$.
\begin{lemma}\label{Lemma2}
The tensor product of two upper triangular matrices is an upper triangular matrix.
\end{lemma}
\textit{Proof:} Let $A, B \in \mathbb{C}^{n\times n}$ are two arbitrary upper triangular matrices. Hence
\begin{align}\label{lemma2-C}
C&=A\otimes B\nonumber\\
&=\left(\begin{array}{ccccc}
a_{11}B&a_{12}B&\cdots&a_{1n}B\\ 
a_{21}B&a_{22}B&\cdots&a_{2n}B\\ 
\vdots&\vdots&\ddots&\vdots\\ 
a_{n1}B&a_{n2}B&\cdots&a_{nn}B\\
\end{array}\right).
\end{align}
Since $A$ is the upper triangular matrix,  $a_{ij}=0,~~\forall i>j$. Ergo
\begin{align}\label{lemma2-C-1}
C=\left(\begin{array}{ccccc}
a_{11}B&a_{12}B&\cdots&\cdots&a_{1n}B\\ 
0&a_{22}B&\cdots&\cdots&a_{2n}B\\
0&0&a_{33}B&\cdots&\cdots\\
\vdots&\vdots&\vdots&\ddots&\vdots\\ 
0&0&\cdots&\cdots&0\\
\end{array}\right).
\end{align}
Due to the fact that $B$ is the upper triangular matrix, each block on the diagonal block of $C$, $a_{ii}B$, is also an upper triangular matrix. Consequently, $C$ is the upper triangular matrix.
$\hfill\blacksquare$\\
\textbf{Corollary 2} Since $A_{1}$ and $A_{2}$ are two upper triangular matrices, $\mathbf{A}_{\mathbf{k}}=A_{k_{1}}\otimes A_{k_{2}}\otimes\cdots\otimes A_{k_{d}}$ is an upper triangular matrix $\forall~\mathbf{k}\in\{{k_{1}, k_{2}, \cdots, k_{d}}\}$.
\begin{lemma}
The first entry (in row $1$ and column $1$) of a tensor product of any matrix with $A_{2}$ (Eq. (\ref{Kraus-amplitudedam-app})) is equal to zero.
\end{lemma}\label{Lemma3}
\textit{Proof:} As the first entry of $A_{2}$, $(a_{2})_{11}=0$, then $C=A_{2}\otimes B\Rightarrow c_{11}=(a_{2})_{11}B=0$.
$\hfill\blacksquare$
\\
\textbf{Corollary 3} The only non-zero first entry of $\mathbf{A}_{\mathbf{k}}$, $(\mathbf{a}_{\mathbf{k}})_{11}\neq 0$, is for the case where 
\begin{align*}
\mathbf{A}_{\mathbf{1}}=A_{1}^{\otimes d}.
\end{align*}
Regarding the diagonal form of $A_{1}$ (Eq. (\ref{Kraus-amplitudedam-app}))
\begin{align}
(\mathbf{a}_{\mathbf{1}})_{11}&=1,\label{11entry}\\
(\mathbf{a}_{\mathbf{1}})_{2^{d}2^{d}}&=(1-\eta)^{\frac{d}{2}},\label{NNentry}\\
(\mathbf{a}_{\mathbf{1}})_{i2^{d}}&=0 ~~~\forall i\neq 1, 2^{d}.\label{iNentry}
\end{align}
Respecting the fact that our system of interest is a $d$-qubit system, one shall adopt the following notation
\begin{align*}
\ket{0}^{\otimes d}=\ket{0,0,\cdots,0}&\equiv\ket{\mathbf{1}},\\
\ket{0,0,\cdots,1}&\equiv\ket{\mathbf{2}},\\
&~~\vdots\\
\ket{1}^{\otimes d}=\ket{1,1,\cdots,1}&\equiv\ket{\mathbf{2}^{d}}.
\end{align*}
In this notation, $\rho _{\text{\tiny{GHZ-like}}}=\ket{\Phi}\bra{\Phi}$ where $\ket{\Phi}=\alpha\ket{\mathbf{1}}+\beta\ket{\mathbf{2}^{d}}$ is the initial probe state. Given the presence of amplitude damping noise, the final is given by
\begin{align}
\sum _{\mathbf{k}=1}^{q^{d}}\mathbf{A}_{\mathbf{k}}\rho _{\text{\tiny{GHZ-like}}}\mathbf{A}_{\mathbf{k}}^{\dagger}&=\sum _{\mathbf{k}=1}^{q^{d}}\left(\sum _{\mathbf{i},\mathbf{j}=1}^{2^{d}}(\mathbf{a}_{\mathbf{k}})_{ij}\ket{\mathbf{i}}\bra{\mathbf{j}}\right)\left(\alpha\alpha^{*}\ket{\mathbf{1}}\bra{\mathbf{1}}+\alpha\beta ^{*}\ket{\mathbf{1}}\bra{\mathbf{2}^{d}}+\alpha ^{*}\beta\ket{\mathbf{2}^{d}}\bra{\mathbf{1}}+\beta\beta ^{*}\ket{\mathbf{2}^{d}}\bra{\mathbf{2}^{d}}\right)\left(\sum _{\mathbf{i}',\mathbf{j}'=1}^{2^{d}}(\mathbf{a}_{\mathbf{k}})_{i'j'}^{*}\ket{\mathbf{j'}}\bra{\mathbf{i'}}\right)\nonumber\\
&=\sum _{\mathbf{k}=1}^{q^{d}}\sum _{\mathbf{i},\mathbf{i'}=1}^{2^{d}}(\mathbf{a}_{\mathbf{k}})_{i1}(\mathbf{a}_{\mathbf{k}})_{i'1}^{*}\alpha\alpha^{*}\ket{\mathbf{i}}\bra{\mathbf{i}'}+(\mathbf{a}_{\mathbf{k}})_{i1}(\mathbf{a}_{\mathbf{k}})_{i'2^{d}}^{*}\alpha\beta^{*}\ket{\mathbf{i}}\bra{\mathbf{i}'}+(\mathbf{a}_{\mathbf{k}})_{i2^{d}}(\mathbf{a}_{\mathbf{k}})_{i'1}^{*}\alpha^{*}\beta\ket{\mathbf{i}}\bra{\mathbf{i}'}+(\mathbf{a}_{\mathbf{k}})_{i2^{d}}(\mathbf{a}_{\mathbf{k}})_{i'2^{d}}^{*}\beta\beta^{*}\ket{\mathbf{i}}\bra{\mathbf{i}'},\label{amplitude-GHZlike}
\end{align} 
where $(\mathbf{a}_{\mathbf{k}})_{ij}$ denotes the $i$th  and the $j$th entry of $\mathbf{A}_{\mathbf{k}}$. Regarding the properties of $\mathbf{A}_{\mathbf{k}}$, one can calculate the each term of Eq. (\ref{amplitude-GHZlike}) separately 
\begin{itemize}
\item if $(\mathbf{a}_{\mathbf{k}})_{i1}\neq 0$: 
      \begin{align}
     (\mathbf{a}_{\mathbf{k}})_{i2^{d}}&= 0~~~~~~~~~~~~~~~~~~~~~(\text{Corollary 1})\label{ai2d},\\
     (\mathbf{a}_{\mathbf{k}})_{i'1}^{*}&=\delta_{i'i}(\mathbf{a}_{\mathbf{k}})_{i1}~~~~~~~(\text{Corollary 1})\label{ai'1}.
      \end{align}
      Moreover, regarding the upper triangular property of $\mathbf{A}_{\mathbf{k}}~(\forall~\mathbf{k})$--- see Corollary 2, $(\mathbf{a}_{\mathbf{k}})_{i1}$ can be non-zero only for $i=1$--- see Corollary 3. Substituting Eqs. (\ref{11entry}), (\ref{iNentry}), (\ref{ai2d}), and (\ref{ai'1}) in Eq. (\ref{amplitude-GHZlike}) yields
      \begin{equation}
      \sum _{\mathbf{k}=1}^{q^{d}}\mathbf{A}_{\mathbf{k}}\rho _{\text{\tiny{GHZ-like}}}\mathbf{A}_{\mathbf{k}}^{\dagger}=\sum _{\mathbf{k}=1}^{q^{d}}\sum _{\mathbf{i}=1}^{2^{d}}(\mathbf{a}_{\mathbf{k}})_{i1}(\mathbf{a}_{\mathbf{k}})_{i1}^{*}\alpha\alpha^{*}\ket{\mathbf{i}}\bra{\mathbf{i}}+(\mathbf{a}_{\mathbf{k}})_{11}(\mathbf{a}_{\mathbf{k}})_{2^{d}2^{d}}^{*}\alpha\beta^{*}\ket{\mathbf{1}}\bra{\mathbf{2}^{d}}.\label{amplitude-GHZlike-1}
\end{equation}

\item if $(\mathbf{a}_{\mathbf{k}})_{i1}=0$: 
      \begin{align}
     (\mathbf{a}_{\mathbf{k}})_{i2^{d}}&\neq 0~~~~~~~~~~~~~~~~~~~~~(\text{Corollary 1})\label{ai2d-2},\\
     (\mathbf{a}_{\mathbf{k}})_{i'2^{d}}^{*}&=\delta_{i'i}(\mathbf{a}_{\mathbf{k}})_{i2^{d}}~~~~~~~(\text{Corollary 1})\label{ai'2d}.
      \end{align}
      Applying a similar method (same as the previous step) to Eq. (\ref{amplitude-GHZlike}) gives
      \begin{equation}
      \sum _{\mathbf{k}=1}^{q^{d}}\mathbf{A}_{\mathbf{k}}\rho _{\text{\tiny{GHZ-like}}}\mathbf{A}_{\mathbf{k}}^{\dagger}=\sum _{\mathbf{k}=1}^{q^{d}}(\mathbf{a}_{\mathbf{k}})_{2^{d}2^{d}}(\mathbf{a}_{\mathbf{k}})_{11}^{*}\alpha^{*}\beta\ket{\mathbf{2}^{d}}\bra{\mathbf{1}}+\sum _{\mathbf{i}=1}^{2^{d}}(\mathbf{a}_{\mathbf{k}})_{i2^{d}}(\mathbf{a}_{\mathbf{k}})_{i2^{d}}^{*}\beta\beta^{*}\ket{\mathbf{i}}\bra{\mathbf{i}}.\label{amplitude-GHZlike-2}
\end{equation}
\end{itemize}
As a consequence
\begin{align}
\sum _{\mathbf{k}=1}^{q^{d}}\mathbf{A}_{\mathbf{k}}\rho _{\text{\tiny{GHZ-like}}}\mathbf{A}_{\mathbf{k}}^{\dagger}&=\sum _{\mathbf{k}=1}^{q^{d}}(\mathbf{a}_{\mathbf{k}})_{11}(\mathbf{a}_{\mathbf{k}})_{2^{d}2^{d}}^{*}\alpha\beta^{*}\ket{\mathbf{1}}\bra{\mathbf{2}^{d}}+(\mathbf{a}_{\mathbf{k}})_{2^{d}2^{d}}(\mathbf{a}_{\mathbf{k}})_{11}^{*}\alpha^{*}\beta\ket{\mathbf{2}^{d}}\bra{\mathbf{1}}+\sum _{\mathbf{i}=1}^{2^{d}}(\mathbf{a}_{\mathbf{k}})_{i1}(\mathbf{a}_{\mathbf{k}})_{i1}^{*}\alpha\alpha^{*}+(\mathbf{a}_{\mathbf{k}})_{i2^{d}}(\mathbf{a}_{\mathbf{k}})_{i2^{d}}^{*}\beta\beta^{*}\ket{\mathbf{i}}\bra{\mathbf{i}}\nonumber\\
&=\rho _{\text{\tiny{GHZ-like}}}(\eta)+\rho_{\text{\tiny{diagonal}}}(\eta).\label{amplitude-GHZlike-total}
\end{align}

%%%%%%%%%%%%%%%%%%%%%%%%%%%%%%%%%%%%%%%%%%%%%%%%%%%%%%%%%%%%%%%%
%\section{}\label{}

%%%%%%%%%%%%%%%%%%%%%%%%%%%%%%%%%%%%%%%%%%%%%%%%%%%%%%

%%%%%%%%%%%%%%%%%%%%%%%%%%%%%%%%%%%%%%%%%%%%%%%%%%%%%%

\bibliography{Ref}
%%%%%%%%%%%%%%%%%%%%%%%%%%%%%%%%%%%%%%%%%%%%%%%%%%%%%%%%%%%%%%%%
\end{document}